\documentclass[useAMS,usenatbib,letterpaper]{mn2e}
\usepackage{times}
\input{epsf}
\title[Multiple outflows in NGC\,6058]
{Multiple outflows in the planetary nebula NGC\,6058}

\author[P. F. Guill\'{e}n et al.]
{
P. F. Guill\'{e}n$^{1}$\thanks{E-mail:fguillen@astro.unam.mx},
R. V\'{a}zquez$^{1}$, L. F. Miranda$^{2,3}$, S. Zavala$^{4,5}$, M. E. Contreras$^{1}$,  \newauthor
S. Ayala$^{6}$ and A. Ortiz-Ambriz$^{7}$\\
$^{1}$Instituto de Astronom\'{\i}a, Universidad Nacional Aut\'onoma de
M\'exico, Apdo. Postal 877, 22800 Ensenada, B. C., Mexico\\
$^{2}$Departamento de F\'{\i}sica Aplicada, Universidade de Vigo, Campus Lagoas-Marcosende s/n, E-36310 Vigo, Spain\\
$^{3}$Consejo Superior de Investigaciones Cient\'{\i}ficas, Serrano 117, E-28006 Madrid, Spain\\
$^{4}$Instituto Tecnol\'ogico de Ensenada, Blvd. Tecnol\'ogico No. 150, 22780 Ensenada, B. C., Mexico\\
$^{5}$Instituto de Estudios Avanzados de Baja California, A. C., Av. Obreg\'on 1755, 22800 Ensenada, B. C., Mexico\\
$^{6}$Facultad de Ciencias F\'{\i}sico-Matem\'aticas, Universidad Aut\'onoma de Nuevo Le\'on, 
Av. Universidad s/n., 66451 San Nicol\'as de los Garza, N. L., Mexico\\
$^{7}$Photonics and Mathematical Optics Group, Tecnol\'{o}gico de Monterrey,
Av. Eugenio Garza Sada. 2501, 64849, Monterrey, N. L., Mexico.\\
}

\begin{document}

\date{Accepted 2013 April 10}

\pagerange{\pageref{firstpage}--\pageref{lastpage}} \pubyear{}

\maketitle

\label{firstpage}
\begin{abstract} 
We present narrow-band [O\,{\sc iii}]$\lambda5007$ and H$\alpha$ images, as well as long-slit 
high-resolution echelle spectra of the planetary nebula NGC\,6058. Our data reveal that NGC\,6058 
is a multipolar planetary nebula of about $\simeq45$\,arcsec in extent and formed by four bipolar 
outflows that are oriented at different position angles. Assuming homologous expansion for all the 
structures, and a distance of 3.5\,kpc, we obtain polar velocities around
$\simeq68\,{\rm km\,s}^{-1}$ for three of them. The estimated kinematical ages suggest that 
the three oldest outflows have been ejected in intervals of $\sim1100$ and $\sim400$\,yr during 
which, the ejection axis has changes its orientation by $\sim60\degr$ and $\sim40\degr$, respectively.
Although a inner ring-like structure is suggested by the direct images, its kinematics shows that no 
equatorial ring or toroid exists in the nebula. At the contrary, the long-slit spectra reveal that the 
ring-like structure corresponds to a fourth outflow that is oriented almost perpendicular to the other 
three. This fourth outflow is the youngest one and appears to be interacting with the other three, 
creating a protruding zone that sweeps material in a region almost perpendicular to the major 
axes of the oldest outflows. This structure also presents two bright arcuate regions along the 
same direction of the older outflows, and at opposite sides from the central star. From our model, 
we suggest that NGC\,6058 could be an intermediate evolutionary stage between starfish 
planetary nebulae and multipolar planetary nebula with apparent equatorial lobes.
\end{abstract}


\section{Introduction}
\begin{figure*}
 \centerline{
\epsfxsize=1.0\textwidth
\epsfbox{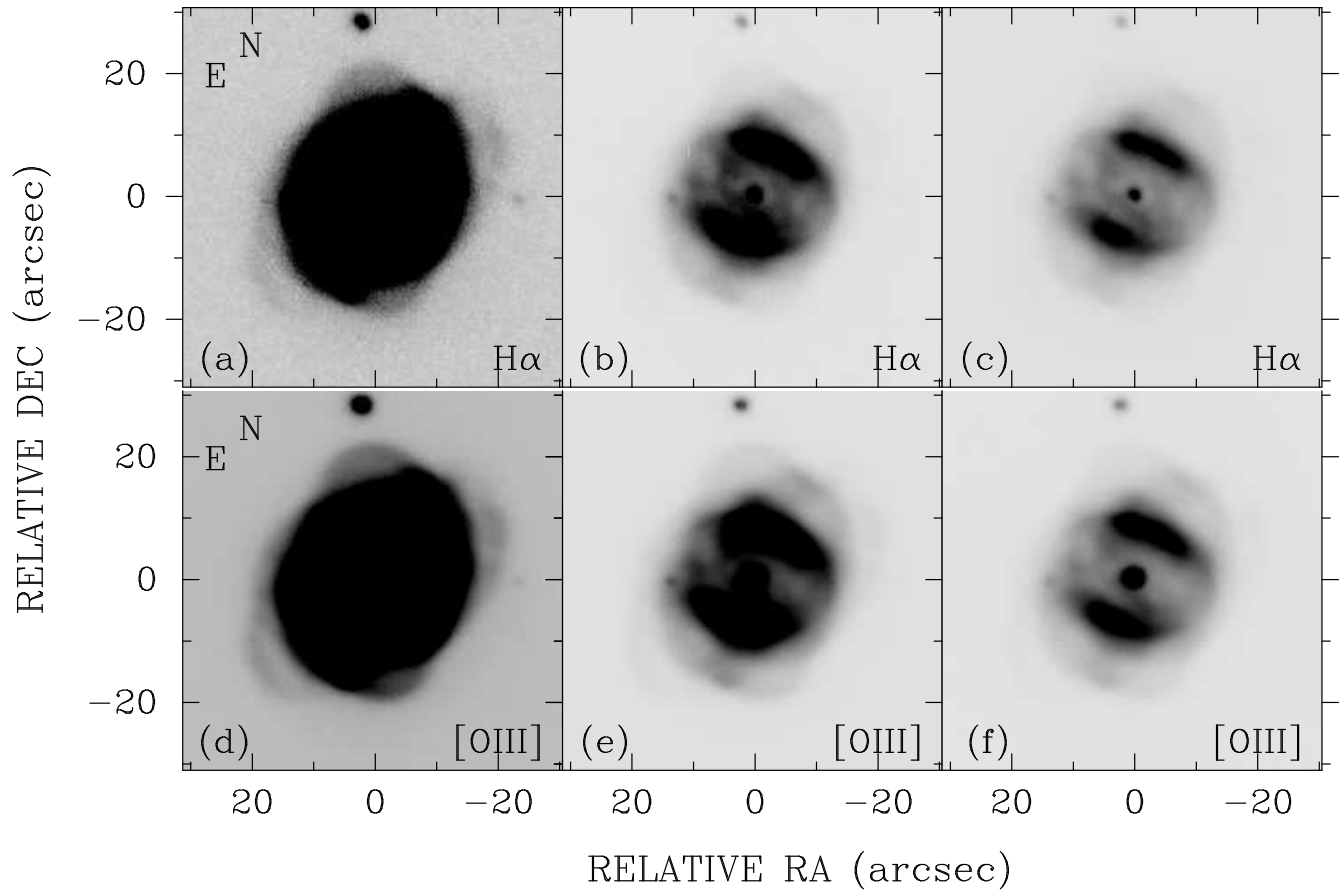}
}
\caption{Greyscale displays of NGC\,6058. Upper (lower) panels show H$\alpha$ ([O\,{\sc iii}])
images with arbitrary contrast chosen to enhance the different features present in the nebula. 
\label{mosaico}}
\end{figure*}

A Planetary Nebula (PN) is a physical system composed by a gaseous shell surrounding an 
evolved star. The shell is a product of the evolution of the star which is often referred as `nucleus' 
or `central star of the PN'. This system corresponds to a particular stage in the evolution of low- 
to medium-mass stars ($M\leq8$\,M${_\odot}$), between the Asymptotic Giant Branch (AGB) and 
the white dwarf phases. The interacting stellar wind model \citep{kwo78} and its generalization 
\citep{bal87,ick89} account for the simplest PN morphologies, namely, spherical, elliptical, and bipolar. 
However, it is well known that PNe show more complex morphologies, including, e.g., jets, multiple 
outflows, and  point-symmetric structures (see, \citealt{ack92,sch92,man96c}). In some cases, 
extremely complex morphologies have been found in images obtained with the Hubble Space 
Telescope\footnote{Many examples of PNe and other objects can be seen at 
http://heritage.stsci.edu/gallery/gallery.html and related sites.}. Thus, in the last decades, a renewed 
interest has risen to explain these complex morphologies. 

Among these complex morphologies, \citet{man96} defined a new morphological class of PNe
they called `quadrupolar', based on the presence of two systems of bipolar lobes in the same PN, 
which present different orientations but share the same center. Examples of quadrupolar PNe 
include K\,3-24, M\,1-75, M\,2-46, M\,3-28, and M\,4-14 \citep{man96}; NGC 6881 \citep{gue98,kwo05}, 
PN\,G126.6+01.3 \citep{mam06}, Kn\,26 \citep{gue13}, and NGC\,6309 \citep{vaz08}. More extreme is 
the case of PNe that show more than two bipolar outflows, the so called multipolar or polypolar PNe 
(e.g. NGC\,2440 \citealt{lop98}, \citealt{vaz99}; He\,2-113, \citealt{sah00b}; J\,320 \citealt{har04}; 
NGC\,5189, \citealt{sab12}). Different explanations have been used in literature to explain the 
presence of more than one system of bipolar lobes, as, for instance, multiple ejections from a binary
star (e.g. \citealt{vel12}), and bipolar ejections collimated by a warped irradiated disk 
(e.g. \citealt{rijk05}). In order to impose constrains on the models for the formation of complex PNe, 
the analysis of their spatio-kinematical structure has proven to be crucial (e.g. 
\citealt{bal87b,mir92,cor93,lop98,mir06,gue08,vaz08,vay09,gar09,con10}). In particular, 
high-resolution, long-slit spectra combined with narrow-band images is a powerful technique to 
obtain the 3D structure of PNe. 

In this work, we analyze the morphokinematical structure of NGC\,6058 (PN\,G064.6+48.2), located at 
$\rmn{RA}(2000)=16^{\rmn{h}} 04^{\rmn{m}} 26\fs5$, $\rmn{Dec.}~(2000)=+40\degr 40\arcmin 56\arcsec$, 
which is a relatively small PN with an apparent elliptical morphology \citep{man96}, 
but also classified as a multiple-shell PN by \citet{chu87}. Based on a kinematical study, 
\citet{sab84} concluded that NGC\,6058 is an evolved PN in the optically thin phase. He reports 
expansion velocities of 26\,km\,s$^{-1}$ in  [O\,{\sc iii}] and 23\,km\,s$^{-1}$ in H$\alpha$. 
NGC\,6058 is a high-excitation PN \citep{gur91} with relatively low electron densities 
($N_{\rm e}$[O\,{\sc ii}]$\simeq1400$\,cm$^{-3}$, and $N_{\rm e}$[Ar\,{\sc iv}] 
$\simeq912$\,cm$^{-3}$, \citealt{wan04}) and electron temperature 
$T_{\rm e}\simeq$13,200$\pm500$\,K \citep{kal85}. Although NGC\,6058 appears in a relatively 
large number of articles (see references in SIMBAD), there is not a detailed analysis of its internal 
kinematics and morphology that allows us to deduce its 3D structure and the physical processes involved 
on its formation. 

\begin{figure*}
 \centerline{
\epsfxsize=7in
\epsfbox{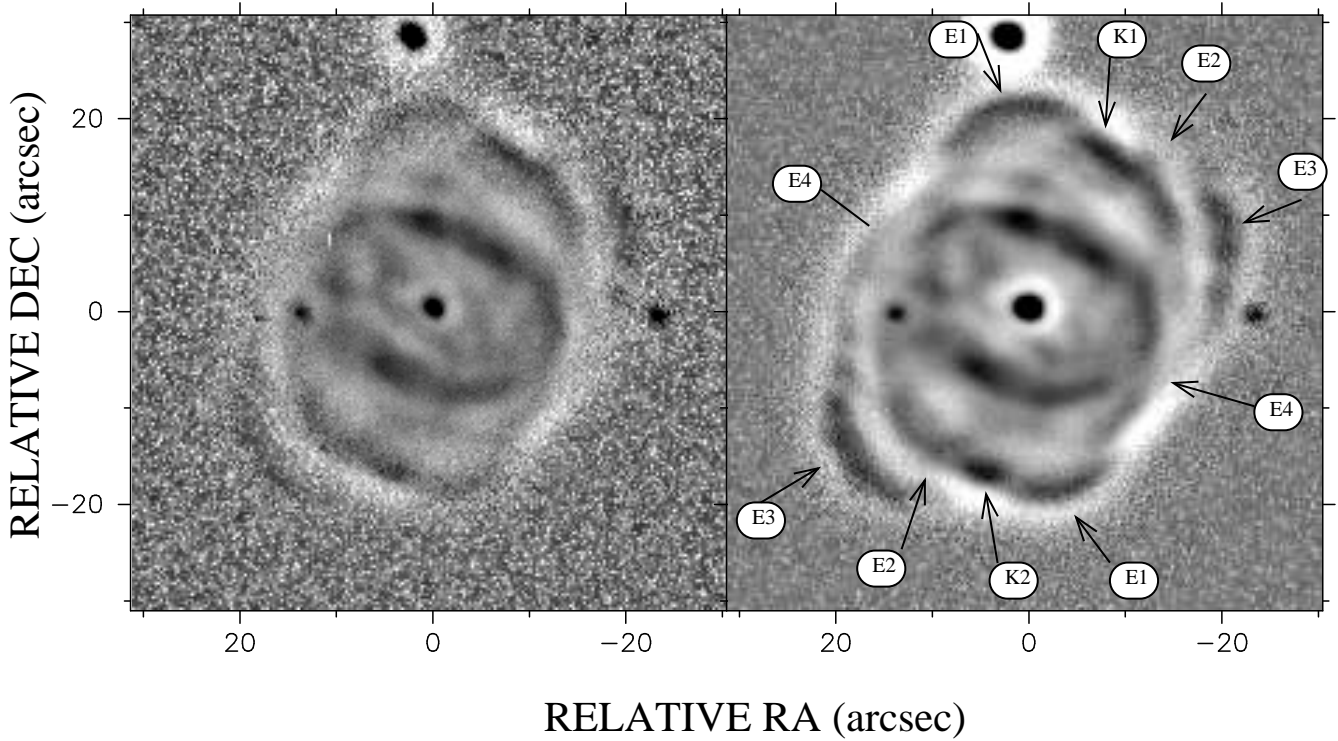}
}
\caption{Greyscale H$\alpha$ (left) and [O\,{\sc iii}] (right) images of NGC\,6058 after being processed 
with an unsharp-masking filter. High spatial frequencies are remarked with this technique. 
The pairs of arcs mentioned in the text are labeled as E1, E2, E3, and E4.
\label{umask2}}
\end{figure*}

In this paper we present a morphokinematical study of NGC\,6058, based on deep imaging and 
high-resolution, long-slit spectroscopy. Our study reveals that NGC\,6058 belongs to the group of 
multipolar PNe mentioned above. In Section\,2 we present our observations and results which 
are discussed in Section\ref{discusion}.

\section{Observations and Results}

\subsection{CCD direct images}
\label{imagenes}

Direct narrow-band images of NGC\,6058 were obtained on 2007 April 18 with the 1.5-m Harold 
Johnson telescope at the San Pedro M\'artir Observatory (OAN-SPM\footnote{The Observatorio 
Astron\'omico Nacional (OAN-SPM) is located at the Sierra de San Pedro M\'artir, Baja California, 
and is operated by the Instituto de Astronom{\'\i}a of the Universidad Nacional Aut\'onoma de M\'exico 
(UNAM).}). The detector was a 2048$\times$2048 Marconi CCD (13.5\,$\mu$m pixel size) with a plate 
scale of 0.14\,arcsec\,pix$^{-1}$. The detector was set to binning 2$\times$2, resulting in a plate scale 
of 0.28\,arcsec\,pix$^{-1}$ and a field of view of 4$\farcm$8$\times$4$\farcm$8. 
H$\alpha$ ($\lambda_c = 6563$\AA, FWHM $= 10$\AA ) and
[O\,{\sc iii}] ($\lambda_c = 5007$\AA, FWHM $= 50$\AA) filters were employed to acquire the images. 
Seeing was around 2 arcsec during observations. Total exposure times were 2100\,s and 3600\,s
for the H$\alpha$ and [O\,{\sc iii}] images, respectively. Images were processed using standard 
techniques of {\sc iraf} and they are shown in Fig.~\ref{mosaico}.

In the H$\alpha$ and [O\,{\sc iii}] images, the nebula resembles an ellipse ($42 \times 48$ arcsec$^2$), 
and recalls the structure previously reported by \citet{chu87} (internal shell), characterized by faint polar 
regions and a bright elliptical ring-like region. However, a detailed inspection of  the images 
(Figure~\ref{mosaico}) reveals the existence of two point-symmetric pairs of faint arcs-like filaments 
beyond the brighter emission (panels a and d), which have not been reported before. 

In order to enhance these structures, we have obtained unsharp-masking images in H$\alpha$ and 
[O\,{\sc iii}] that are shown in Fig.~\ref{umask2}. These images reveal three pairs of point-symmetric 
arc-like filaments, labelled E1, E2, and E3, and the ring-like region, labelled E4. The pairs E1 and E3 
correspond to the new point-symmetric arcs detected in the H$\alpha$ and [O\,{\sc iii}] images, while 
E2 seems to trace the polar caps of the elliptical shell found by \citet{chu87}. Thus, E1, 
E2, and E3 seem to be the polar caps of ellipses whose semi-major axes have different positions angles. 
The pair E1 is oriented at $\rmn{PA}\simeq7\degr$ and presents an extent of $\simeq38$\,arcsec. E2 is 
oriented at $\rmn{PA}\simeq-32\degr$ and its extent is $\simeq33$\,arcsec. Between E1 and E2 two enhanced emission regions (denoted K1 and K2, Fig.~\ref{umask2}) can be distinguished at opposite 
sides of  the  central star. The pair E3 is oriented at $\rmn{PA}\simeq-52\degr$ and presents an 
extent of $\simeq44$\,arcsec. The ring-like structure E4 presents a size of $\simeq16$\,arcsec $\times$ 
27~arcsec, with the major axis oriented at $\rmn{PA}\simeq65\degr$. Contrary to the other pairs, 
the polar caps of E4 are only marginally detected, whereas the regions around its minor axis are the 
brightest ones of the nebula. 

The images strongly suggest that NGC\,6058 is a multipolar PN containing three bipolar outflows at 
different orientations and a equatorial ring-like structure that is tilted with respect to the line of sight. 
In some respects, the new images of NGC\,6058 show a structure that is reminiscent of starfish PNe 
\citep{sah00}, that present multipolar lobes accompanied by an equatorial ring-like structure. 
However, as we will see below, the kinematics of NGC\,6058 shows that important projection effects 
are involved in this PN and that E4, rather than a equatorial ring, corresponds to a limb-brightening 
barrel-like shell, similar to that seen in the H$\alpha$ image of NGC\,7354 \citep{con10}. 

\begin{figure}
 \centerline{
\epsfxsize=3in
\epsfbox{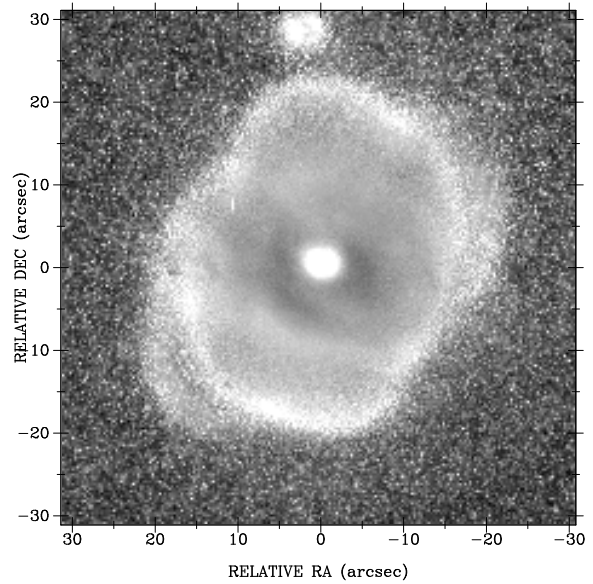}
}
\caption{Greyscale image of the [O III]/H$\alpha$ ratio of NGC\,6058. High and low values of this 
ratio are in white and dark gray, respectively.
\label{cociente}}
\end{figure}

We have also obtained an [O\,{\sc iii}]/H$\alpha$ image ratio that is presented in Fig.~\ref{cociente}. 
This image shows enhancements of the [O\,{\sc iii}] emission at the arcs E1, E2, and E3. According 
to \citet{med09}, the [O\,{\sc iii}] enhancement indicates the presence of shocks at the locations of 
the arcs, which suggests that E1, E2, and E3 are related to collimated outflows.

\subsection{Long-slit echelle spectroscopy}

High-resolution, long-slit spectra were obtained on 2003 June 6 with the Manchester Echelle 
Spectrograph (MES; \citealt{mea03}) in the 2.1-m telescope at OAN-SPM. A Site 24-$\mu$m\,pix$^{-1}$ 
CCD with $1024\times1024$ pixels was used as detector in $2\times2$ binning mode 
(0.6 arcsec\,pix$^{-1}$ plate scale). The spectra were centered at the H$\alpha$ emission line using a 
filter ($\Delta\lambda$=90\AA) to isolate the 87$^{\rm th}$ order (0.1 {\AA}\,pix$^{-1}$  spectral scale). 
Exposure time was 900\,s for each spectrum. 

A second set of spectra were obtained on 2007 July 17 and 18 using the same telescope, instrument 
and detector described above. In this case, the spectra were centered at the [O\,{\sc iii}]$\lambda5007$ 
emission line using a filter ($\Delta\lambda$=50\AA) to isolate the 114$^{\rm th}$ order 
(0.08 {\AA}\,pix$^{-1}$  spectral scale). In this case, exposure time was 1800\,s for each spectrum.

Data were calibrated using standard techniques for long-slit spectroscopy of {\sc iraf}. The slit width
was 150\,$\mu$m (1.9\,arcsec). The resulting spectral resolution (FWHM) is $\simeq12\,{\rm km\,s}^{-1}$ 
(accuracy $\pm1\,{\rm km\,s}^{-1}$), as measured from the lines of the ThAr calibration lamp.  
Seeing was around 1.5~arcsec during the observations.

\begin{figure}
 \centerline{
\epsfxsize=3in
\epsfbox{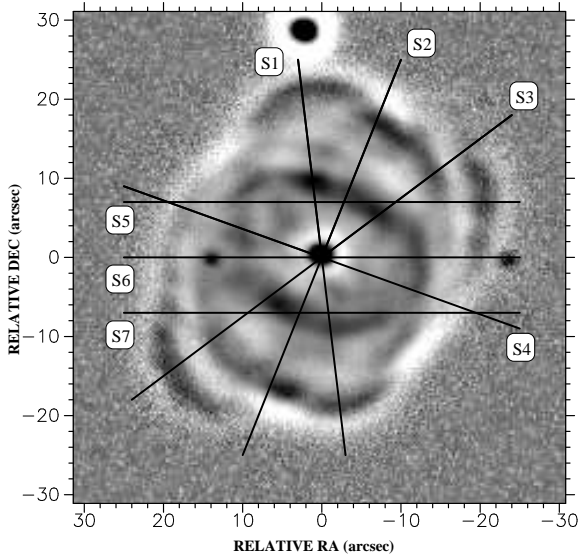}
}
\caption{Unsharp-masked image of NGC\,6058 in [O\,{\sc iii}]. Lines indicate the different slit positions 
that were used to obtain our high-dispersion spectra. Slits labeled as S1, S2, S3, and S4, correspond 
to the spectra centered on [O\,{\sc iii}]$\lambda5007$, whereas slits labeled as S5, S6, and S7, 
correspond to the spectra centered on H$\alpha$.
\label{slitpos}}
\end{figure}

\begin{figure*}
\epsfbox{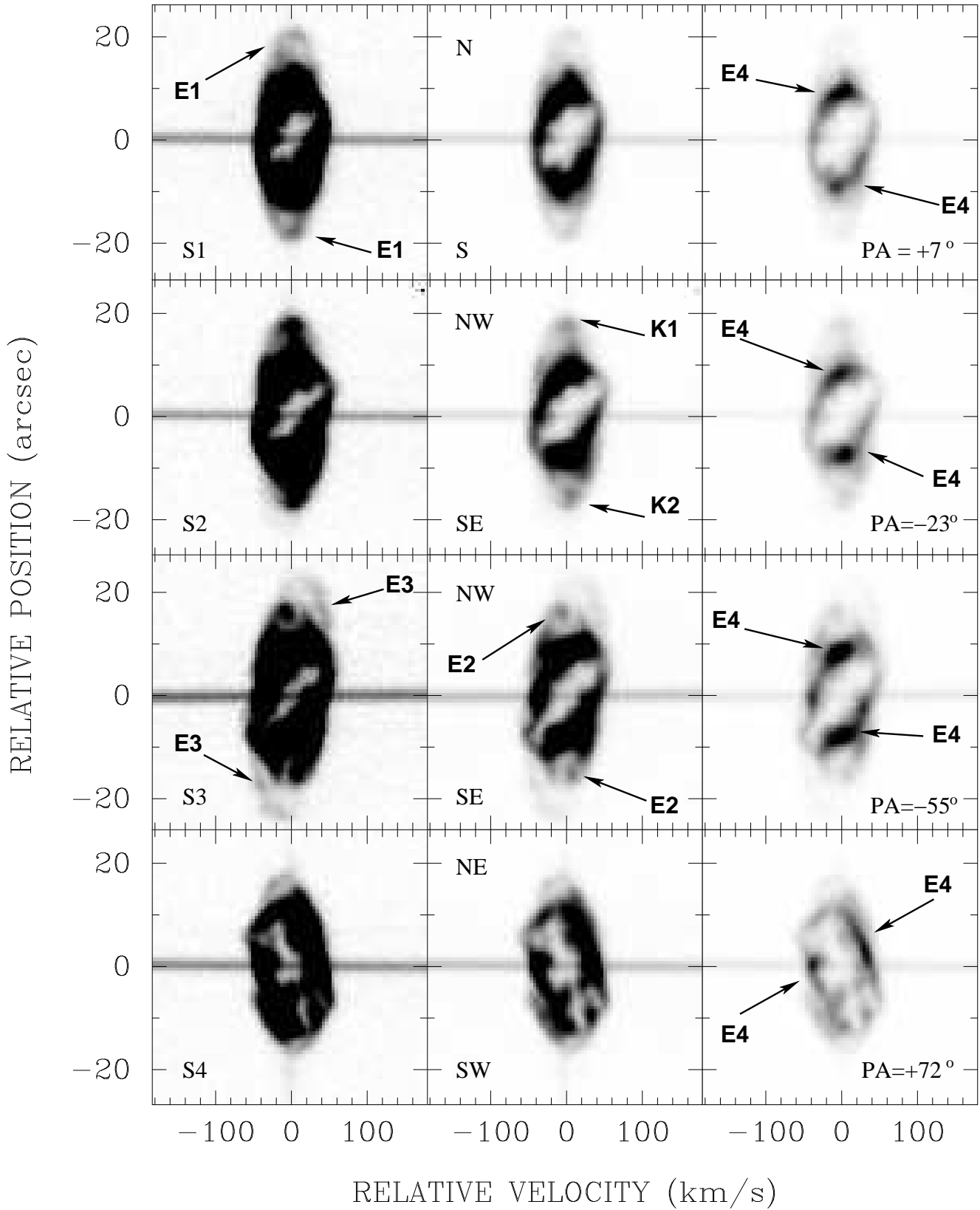}
 \caption{Position-Velocity (PV) maps of the [O\,{\sc iii}]$\lambda5007$ emission line corresponding to 
the S1, S2, S3, and S4 slit positions (rows). Three different grey levels (columns) were arbitrary chosen 
in order to highlight different features in the maps. Kinematical features corresponding to E1, E2, E3, and E4
are labelled in the figure.
\label{specO3}}
\end{figure*}

\begin{figure}
 \centerline{
\epsfxsize=3in
\epsfbox{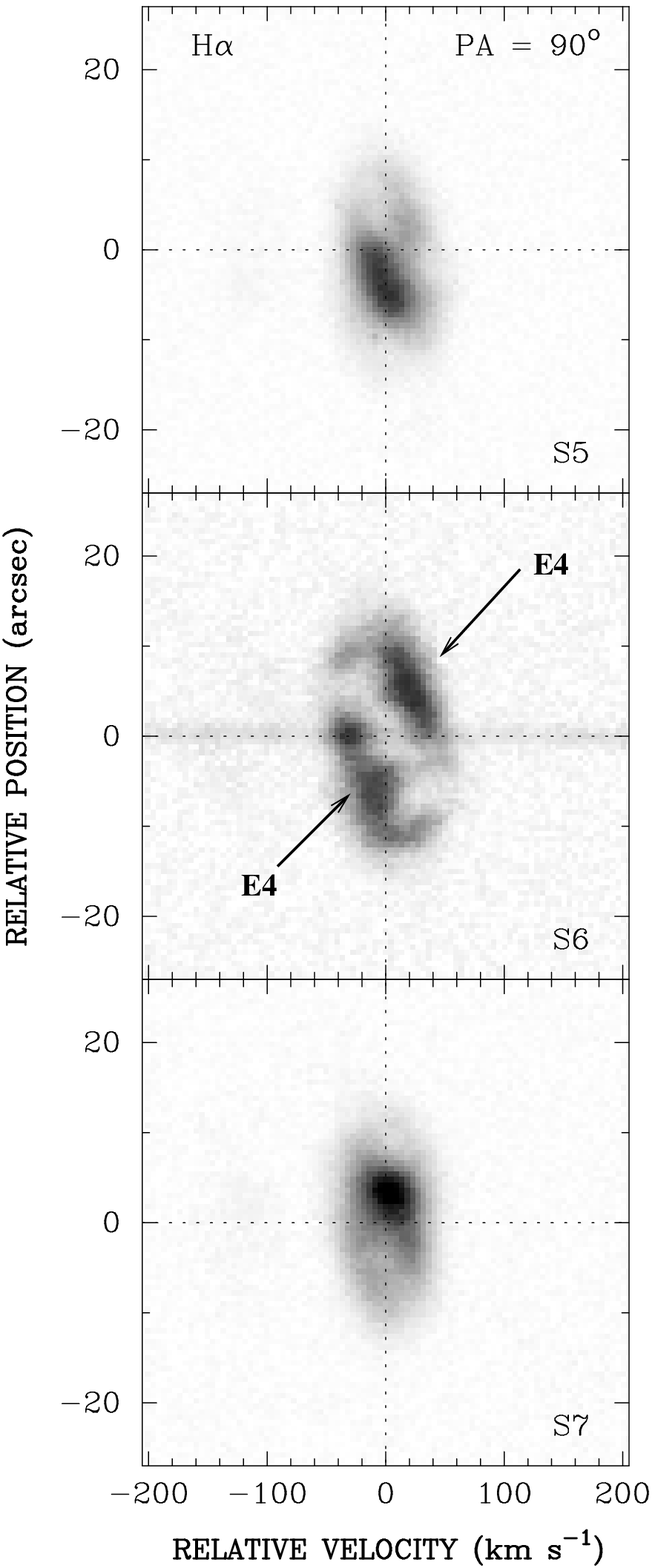}
}
\caption{Position-Velocity maps in H$\alpha$. 
Labels correspond to slit position shown in Fig.~\ref{slitpos}. East is up and West is down.
\label{specHa}}
\end{figure}

Figure~\ref{slitpos} shows the slit positions used to acquire the spectra, superimposed on the [O\,{\sc iii}] 
unsharp-masking image and labeled as follows: S1 ($\rmn{PA} = 7\degr$), S2 ($\rmn{PA} = -23\degr$), 
S3 ($\rmn{PA} = -55\degr$), and S4 ($\rmn{PA} = 72\degr$), for the spectra taken in [O\,{\sc iii}] (all of 
them crossing the central star), and S5, S6, and S7 for the spectra taken in H$\alpha$, all of them 
oriented at ${\rm PA}=90\degr$ with S6 crossing the central star and S5 and S7
shifted 7~arcsec towards the North and the South, respectively.
The corresponding Position-Velocity (PV) maps are shown in 
Figures~\ref{specO3} and \ref{specHa} for [O\,{\sc iii}] and H$\alpha$, respectively. In 
Figure~\ref{specO3}, rows in the PV maps array correspond to the slit positions S1, S2, S3, and S4. 
Columns 1 to 3 show the same spectrum with different contrast levels. Velocity axes of all the PV maps 
are relative to the systemic velocity whereas spatial axes are relative to the location of the central star,
or its Right Ascension, in the cases of S5 and S7. The systemic velocity was obtained from the velocity 
splitting of the spectral lines observed at the position of the central star. We obtain 
$V_{\rm LSR} \simeq 23\pm\,1{\rm km\,s}^{-1}$ ($V_{\rm Hel} \simeq 6\pm1\,{\rm km\,s}^{-1}$), in good agreement with the value of $\simeq20\pm3.3\,{\rm km\,s}^{-1}$ reported by \citet{sch83}. In addition,
the mean value of the expansion velocity, measured at the position of the central star, is 
$v_{{\rm c}x}=37\,{\rm km\,s}^{-1}$.

On the PV maps (Figure~\ref{specO3}), the arc-like structures E1, E2, and E3 identified in the images 
(Fig.~\ref{umask2}; right) have a corresponding structure in the spectra. Each pair appear as the tips of 
an apparent velocity ellipse. In the case of E1, emission can be identified at S1 with an extent of 
$\pm19$~arcsec from the central star. The radial velocities at the tips of E1 differ by 
$7.6\,{\rm km\,s}^{-1}$, with the northern region being redshifted and the southern region blueshifted. 
E2 can be identified at S2 and S3 at $\pm16.4$~arcsec from the central star. The northwestern tip of 
E2 is blueshifted by $4.2\,{\rm km\,s}^{-1}$ while the southeastern tip is redshifted by the same amount. 
We note the existence of two bright knots on the PV map separated by $\simeq33$~arcsec and at the 
systemic velocity, which correspond to the knots K1 and K2 (see Fig.~\ref{umask2}).
Finally, E3 is identified at S3 as two extended features with a spatial extent of $\pm23$~arcsec and a 
difference of radial velocities of $36\,{\rm km\,s}^{-1}$. The northwestern region of E3 redshifted
whereas its southeastern region is blueshifted. 

The PV maps at S1, S2, and S3 also show bright ``knots''  located close to the systemic velocity. 
These bright ``knots'' corresponds to the bright arcs of the ellipse identified in the image as E4. 
Remarkably, the radial velocity of these features is very similar to the systemic velocity. This indicates 
that the bright arcs should move mainly in the plane of the sky and, therefore, it rules out that E4
corresponds to a tilted ring-like structure. Moreover, the bright ``knots'' appear on the PV maps as a 
part of a velocity ellipse, indicating that they belong to a three dimensional  structure. The velocity 
ellipse is well defined at S1 and S2 but presents strong distortions at S3 and S4. In particular at S4, 
the velocity ellipse seems to be open and its apparent polar regions present a complex kinematics. 
These regions correspond to the polar regions of E4 in the images, which are noticeably fainter than 
the arcs. 

Thus, the PV maps reveal the existence of a distorted probably ellipsoidal structure that can be only 
identified in the PV maps but not in the images. This structure, which we continue labeling E4, is tilted 
in such a way that its NE part is blueshifted and its SW part is redshifted. Maximum radial velocities of 
$\simeq\pm53$\,km\,s$^{-1}$ are observed at S4 where a maximum angular size of 24~arcsec is also observed. This implies that the major axis of E4 should be oriented close to PA~72\degr, where the 
maximum distortions are also observed. This orientation is almost perpendicular to that of the 
point-symmetric arcs E1, E2, E3.

The H$\alpha$ emission feature at S6 (Fig.~\ref{specHa}) shows a structure very similar to that 
observed in [O\,{\sc iii}] at S4 (Fig.~\ref{specO3}). In particular, a tilted velocity ellipse is observed with the 
eastern region blueshifted and the western region redshifted. In addition, a marginal evidence of the 
expansion of the East and West sides of E2 is observed in S5 and S7 slit positions, respectively 
($V_{\rm exp}\simeq30$\,km\,s$^{-1}$). 

A final note about the spectra around H$\alpha$ is that the high-excitation nature
of NGC\,6058 results evident in them, given that there is not any detection of 
[N\,{\sc ii}]$\lambda6548,6583$ emission lines in these spectra.
On the contrary, the high-excitation He\,{\sc ii}$\lambda6560$ emission line is 
marginally detected in the spectra, with an intensity of $\leq3\%$  that of H$\alpha$. 
This implies that the contribution of He\,{\sc ii}$\lambda6560$ to the H$\alpha$ image 
(Fig.~\ref{mosaico}) can be considered negligible.

\section{Discussion}  
\label{discusion}

From the morphological and kinematical data mentioned in the last section,
we have identified three bipolar outflows (E1, E2, and E3) whose small differences
between the radial velocities of their polar caps suggest that these outflows were ejected
almost in the plane of the sky. A fourth structure (E4), that easily can be misidentified  
as an equatorial ring in the images, is clearly recognized as an ellipsoidal-like shell in 
the PV maps. The corresponding features of this ellipsoid in the PV maps show a higher 
difference in its maximum radial velocities, suggesting that its orientation is very different 
with respect the outflows E1, E2, and E3.

\begin{figure}
\centerline{
\epsfxsize=2.8in
\epsfbox{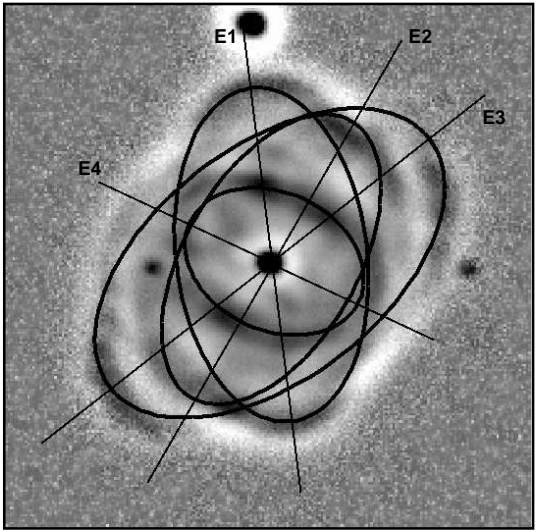}
}
\caption{We present the four ellipses model that best fit the morphology observed in NGC,6058. 
The ellipses are superimposed on the  [O\,{\sc iii}]$\lambda5007$ unsharp-masking image shown
in Fig.~\ref{umask2}{\it b}.
\label{elipses}}
\end{figure}

Based on the apparent elliptical shape of the outflows in both, images and PV maps, 
we have modeled the structures E1, E2, E3, and E4 as ellipsoids. As expected for
these structures, we have assumed that each ellipsoid projects an ellipse in the plane
of the sky which is fitted using {\sc iraf+ds9} software. The ellipses were obtained trying 
to get the best size and shape fits for the polar caps E1, E2, and E3, and for the bright
arcs of E4. Figure~\ref{elipses} shows the fits for the case of the direct image. Using these 
ellipses as the projection in 3D of the ellipsoids, we fit other parameters (inclination angle, 
velocity field, etc.) in order to reproduce the PV maps.

\begin{figure}
\centerline{
\epsfxsize=0.75\columnwidth
\epsfbox{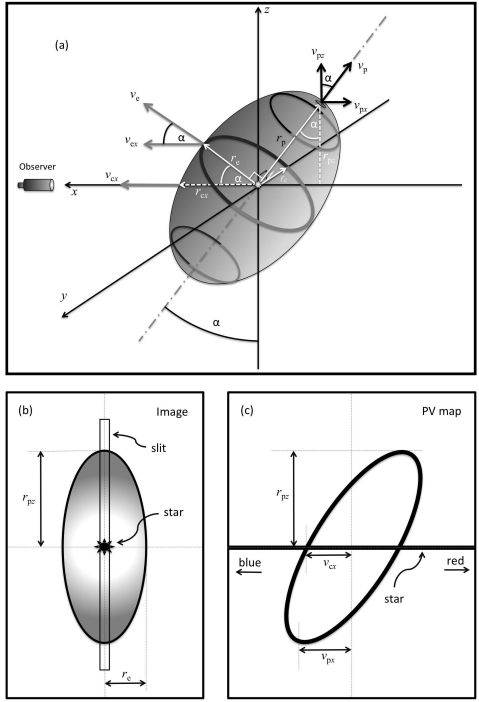}
}
\caption{(a) A sketch of an expanding ellipsoidal structure, whose semi-major axis has an inclination 
angle $\alpha$ respect to the plane of the sky, seen from an observer on Earth. 
Main physical parameters are labeled. (b) The ellipse (image) seen by the observer. 
(c) The position-velocity map (spectrum) from a slit crossing the major axis.}
\label{sketch}
\end{figure}

For each ellipsoid, the apparent polar radius and velocity are measured from the image and the 
corresponding long-slit spectrum, whereas the equatorial radius and velocity are obtained from the
fitted ellipse and the kinematical model, respectively. For the model we have always assumed a 
homologous expansion velocity ($v\propto r$) and small deviations ($\la20\degr$) around 
$90\degr$ for the inclination angle ($i$). A sketch showing the main parameters of an ellipsoidal 
shell (a), as well as the observer view for the image (b) and the PV map (c), are presented in 
Figure~\ref{sketch}. With the considerations mentioned above, and following the nomenclature in
Figure~\ref{sketch}(a), it can be demonstrated that 

$$\tan\,\alpha=\Bigl({r_{\rm e}\over r_{{\rm p}z}}\Bigr)\,\Bigl({v_{{\rm p}x}\over v_{\rm e}}\Bigr)$$

\noindent where $r_{\rm e}$ is the equatorial radius (semi-minor axis); 
$r_{{\rm p}z}$ is the apparent polar radius (projected semi-major axis); 
$v_{\rm e}$ is the equatorial velocity, and 
$v_{{\rm p}x}$ is the apparent polar velocity (radial component).
The inclination angle of the ellipsoid major axis, with respect to the line-of-sight, $i$, can be
derived as $i=90\degr+\alpha$, if the North pole of the ellipsoid is moving away from the
observer, and as $i=90\degr-\alpha$ if coming closer.

It is important to mention that, as can be noted in Fig~\ref{sketch}, the radial velocity
measured at the position of the central star, $v_{{\rm c}x}$, is different than $v_{{\rm e}}$
as well as $v_{{\rm e}x}$. However, using a rough approximation of the polar-to-equatorial 
ratio $r_{\rm p}/r_{\rm e}\approx2$ (which is reasonable, given the semi-axes ratios of the 
ellipses), assuming a ``small'' value for $\alpha\approx20\degr$, and using the cartesian 
equation for a vertical ellipse, we have estimated the ratio

$${v_{{\rm c}x}\over v_{\rm e}}={r_{{\rm c}x}\over r_{\rm e}}=
{{r_{\rm p}/r_{\rm e}}\over\sqrt{\sin^2\alpha+(r_{\rm p}/r_{\rm e})^2\cos^2\alpha}}\approx1.047$$

\noindent i.e., an error $\la5\%$ is commited when considering the approximation 
$v_{\rm e}\approx v_{{\rm c}x}$. Therefore, we will use this approximation and
its associated error to estimate $\alpha$, $i$, and the other parameters in the following calculations.  

Polar radius and velocity can be now obtained using the relationships for the projected $r_{\rm p}$
and the homologous velocity law, respectively

$$r_{\rm p}={r_{{\rm p}z} \over \sin\,i}; \phantom{XX} 
v_{\rm p}=v_{\rm e}\,\Bigl({r_{\rm p}\over r_{\rm e}}\Bigr)=
v_{{\rm c}x}\,\Bigl({r_{{\rm p}z}\over r_{\rm e}\sin\,i}\Bigr).$$

In addition, the kinematical age $\tau_{\rm k}$ can be derived as

$$\Bigl[ {\tau_{\rm k} \over {\rm yr}} \Bigr] = 4744 \times
{{ \Bigl[ {r_{\rm p} \over {\rm arcsec}} \Bigr] 
\Bigl[ {D \over {\rm kpc}} \Bigr] }
\Bigl[ {v_{\rm p} \over {\rm km\,s}^{-1}} \Bigr]^{-1}} $$

\noindent where $D$ is the distance from the Earth to the nebula (in kpc). 
Values of $i$, $r_{\rm p}$, $v_{\rm p}$, and $\tau_{\rm k}$, obtained for
all the ellipsoids and assuming a distance of 3.5\,kpc \citep{cah92}, are 
presented in Table~\ref{elipso}. It is important to mention that E4 was 
excluded from Table~\ref{elipso}, given that is not possible to obtain a suitable kinematic 
model for this ellipse. As we will explain below, we interpret this as the result of 
outflows interaction. However, as a reference only, we propose the following limits for this ratio

$$1.3 \la {r_{\rm p}\over r_{\rm e}} \la 2.$$

\noindent where the lower limit corresponds to the apparent axes ratio ($r_{{\rm p}z}/ r_{\rm e}$)
seen in E4, and the upper limit was choose as a reasonable value given the apparent polar-to-equatorial
ratios from the other ellipses. Using these values, the possible inclination angle for such ellipse 
would be between 50 and $130\degr$. Consequently, a kinematical age for E4 can be estimated
as $\tau_{\rm k}\approx3400$\,yr. However, as we mentioned above, this value must be taken 
cautiously.

\begin{table*}
 \caption{Parameters of ellipsoids measured and derived from models: 
 position angle (PA), semi-minor axis ($r_{\rm e}$), apparent semi-major axis ($r_{\rm p}$),
 equatorial and polar apparent velocities ($v_{\rm e}$) and ($v_{\rm p}$), inclination angle ($i$), 
 semi-major axis ($r_{\rm p}$), polar velocity ($v_{\rm p}$), and kinematical age ($\tau_{\rm k}$).}
\begin{tabular}{@{}cccccccccc@{}}
\hline
\hline
      &    PA     &  $r_{\rm e}$ & $r_{{\rm p}z}$ & $v_{{\rm c}x}$ & $v_{{\rm p}x}$ 
      &  $i$    &  $r_{\rm p}$ & $v_{\rm p}$ & $\tau_{\rm k}$\\
      &  (\degr) &  (arcsec)    & (arcsec)        & (km\,s$^{-1}$)  & (km\,s$^{-1}$)
      & (\degr) &  (arcsec)    & (km\,s$^{-1}$)      & (yr)          \\
\hline 
 E1   & +7      & $11\pm1$ & $19\pm1$ & $37\pm2$ & $3.8\pm1$ &  $93\pm2$ & $19\pm1$  
 & $67\pm8$  & $4\,800\pm600$      \\
 E2   & $-30$ & $10\pm1$ & $18\pm1$ & $37\pm2$ & $4.2\pm1$ &  $86\pm2$ & $18\pm1$ 
 & $69\pm9$  & $4\,400\pm600$      \\
 E3   & $-52$ & $13\pm1$ & $23\pm1$ & $37\pm2$ & $18\pm1$ &$106\pm2$ & $24\pm1$    
 & $67\pm10$  & $5\,900\pm600$      \\
\hline
\end{tabular}
\label{elipso}
\end{table*}

As can be noted in Table~\ref{elipso}, given the uncertainties in the kinematical ages, it is
not simple to describe the sequence of events that have occurred in the formation of NGC\,6058. 
Nevertheless, based on the kinematical ages as well as on the orientation of the structures, 
we propose a scenario in which a series of bipolar outflows could shape the morphology of 
NGC\,6058. In our proposal, the first outflow was E3 along ${\rm PA}\simeq-52\degr$. 
The second ejection was E1 along ${\rm PA}\simeq7\degr$, involving a change in the 
orientation of the ejection axis of $\simeq60\degr$ with respect to that of E1. The third ejection 
was E2 along ${\rm PA}\simeq-30\degr$ that involves a change of orientation of $\simeq40\degr$ 
with respect to E1. We note that E3 and E1 are ejected with their northern/northwestern half 
pointing away from the observer whereas the northwestern half of E2 points towards the observer. 

Moreover, the deduced kinematical ages (Table~\ref{elipso}) suggest a time span of $\sim1100$~yr 
between the two first collimated ejections and only $\sim400$\,yr between the second and the third. 
These results point to the existence of episodic ejections with changes in the orientation of the main
axis, suggesting precession of the collimation axis. About $\ga1000$~yr after ejection of E2, a new
shell is formed which is identified with E4. In this case, the main ejection axis has largely changed 
with respect to that of E1--E3, as the polar axis of E4 is almost perpendicular to the previous outflows. 
Our data strongly suggest that E4 interacts with the previous outflows. The bright arc-like regions 
observed in the direct images are probably a result of the interaction of the equatorial regions of 
E4 with parts of E1--E3. Furthermore, the large kinematical distortions observed in E4 along PA 
around $72\degr$ suggest that E4 also interacts with the equatorial regions of the previous outflows 
and protrudes. We note that this interaction may result in a deceleration of E4 and, therefore, the
deduced kinematical age of 3400\,yr is probably an upper limit, implying that E4 has been formed 
noticeably later than the previous outflows.

As a result of this study we can conclude that NGC\,6058 belongs to the group of multipolar PNe. 
It is interesting to compare NGC\,6058 with other multipolar PNe. In particular, we find remarkable 
similarities between NGC\,6058 and the so called `starfish PNe' \citep{sah00}. These PNe present 
three or four pairs of bipolar lobes and an bright elliptical ring-like structure that can be interpreted 
as a tilted equatorial ring. We may relate the E1, E2, and E3 outflows with the multiple pairs of lobes 
of starfish PNe, and the bright arcs corresponding to E4 with the equatorial ring of the starfish. 
However, in NGC\,6058 our kinematical data reveal that the bright arcs do not trace a tilted equatorial 
ring but a part of an ellipsoid shell. In this respect, it would be interesting to study the internal 
kinematical of starfish PNe in order to check whether the observed elliptical rings have the kinematics 
expected from a tilted ring or, at the contrary, they trace an ellipsoidal shell, as in the case of NGC\,6058.

It is interesting to speculate about the possible future evolution of NGC\,6058. In particular, the polar 
regions of E4 could break and expand at a higher velocity, creating, after some time, a pair of large 
bipolar structures. If so, NGC\,6058 would consist in a pair of extended bipolar structures, 
accompanied of smaller bipolar lobes around its equatorial region, which would correspond to 
E1--E3. This scenario could provide an explanation for some PNe like Hb\,5 and K\,3-17 
\citep{mir10,lop12} in which large bipolar lobes are observed accompanied by secondary smaller
lobes around the equator of the largest ones. By all these comments, we suggest that NGC\,6058 
could be an intermediate evolutionary stage between starfish PNe and PNe with equatorial bipolar lobes.

\section{Conclusions}

We have carried out a morphokinematic analysis of the planetary nebula NGC\,6058, based on direct 
imaging and high-dispersion spectroscopy. NGC\,6058 can be described as a multipolar PN, formed by 
three bipolar outflows with different orientations and polar velocities around $\sim68\,$km\,s$^{-1}$, as
well as a distorted ellipsoidal inner shell with its major axis almost perpendicular to those of the other three 
outflows. Assuming homologous expansion for all the structures and a distance of 3.5\,kpc, we obtain 
kinematical ages ranged from $\sim3400$ to $\sim5900$\,yr for the different outflows.
Our data suggest a certain systematic behaviour in the changes of the orientation of the main
axis, which points out to episodic, precessing ejections in NGC\,6058. The ellipsoidal inner shell is
released much after than the three collimated outflows and its kinematics reveals clear signs of interaction 
with the previous ejections. A comparison with other multipolar PNe suggests that NGC\,6058 could 
be an intermediate stage between (young) starfish PNe and more evolved PNe that show a large pair 
of bipolar lobes and much smaller equatorial lobes.

\section*{Acknowledgments}

We are grateful to the staff of OAN-SPM, specially to Mr. Gustavo Melgoza-Kennedy,
our telescope operator, for his assistantship during observations. This paper has been supported by 
Mexican grant IN109509 (PAPIIT-DGAPA-UNAM).
LFM is partly supported by grant AYA2011-30228-C03.01 of the Spanish Ministerio 
de Econom\'{\i}a y Competitividad (MINECO) and by grant IN845B-2010/061 of Xunta de Galicia
(Spain), all of them partially funded by FEDER funds. SZ acknowledges support from the UNAM-ITE 
collaboration agreement 1500-479-3-V-04. SA acknowledges support from the UNAM-UANL
collaboration agreement. This research has made use of the SIMBAD database,
operated at CDS, Strasbourg, France. IRAF is distributed by the National Optical Astronomy 
Observatories, which are operated by the Association of Universities for Research in Astronomy, 
Inc., under cooperative agreement with the National Science Foundation. We thank our anonymous 
referee for valuable comments which have improved this article. Authors want to dedicate this paper 
in memorial of Gaby Garc\'{\i}a-Ruiz, member of the OAN-SPM staff, who passed away.\\

\label{lastpage}

\end{document}